\documentstyle[pre,aps,psfig]{revtex}
\begin{document}




\draft

\title{Long range correlations in DNA sequences}
\author{A. K. Mohanty and A. V. S. S. Narayana Rao$^*$}
\address{Nuclear Physics Division, Bhabha Atomic Research Centre, Mumbai-400085}
\address{$^*$Molecular Biology and Agriculture Division,Bhabha Atomic Research Centre, Mumbai-400085}

\maketitle

\begin{abstract}
The so called long range correlation properties of DNA
sequences are studied using the  variance analyses of the density
distribution of a single or a group of nucleotides in a model independent way.
This new method which was suggested earlier has been applied to  extract
slope parameters that characterize the correlation properties for several
intron containing and intron less DNA sequences. An important aspect of all the DNA
sequences is the properties of complimentarity by virtue
of which any two complimentary
distributions (like $GA$ is complimentary to $TC$ or $G$ is complimentary to $ATC$)
have identical fluctuations at all scales although their distribution functions
need not be identical. Due to this complimentarity, the famous DNA walk
representation whose statistical interpretation is still unresolved is shown
to be a special case of the present formalism with a density distribution
corresponding to a purine or a pyrimidine group. Another interesting aspect
of most of the DNA sequences is that the factorial moments
as a function of length exceed unity around a region where the variance
versus length in a log-log plot shows a bending. This is a pure phenomenological
observation
which is found for several DNA sequences with a few exception. Therefore, this
length scale has been used as an approximate measure to exclude the bending regions
from the slope analyses. The asymmetries in the nucleotide contents or
the patchy structure as a possible origin of the long range correlations has
also been investigated.

\end{abstract}


\section{INTRODUCTION}
Recently, there has been considerable interest in the finding of long range
correlations in genomic DNA sequences \cite{LI1}. A DNA sequence is a chain
of sites, each occupied by either a purine (Adenine and Guanine) or a
pyrimidine (Cytocyine and Thymine) group. For mathematical modeling, the DNA
sequence might be considered as a string of symbols (G, A, T and C) whose
correlation structure can be characterized  completely by all possible base-base
correlation functions or their corresponding power spectra. Different techniques
including mutual information functions and power spectra analyses
\cite{LI1,LI2,LI3,LI4,VOSS,BUL1,BOR,LU,VIE}, auto correlation \cite{AZB,HER,LUO}, DNA
walk representation \cite{PENG1,MAD,NEE,CHA,PRA,KAR,STA,BUL2}, wavelet
analysis \cite{ARN1,ARN2} and Zipf analysis \cite{MAN} were used for statistical
analyses of DNA sequences. But despite the effort spent, it is still an
open question whether
the long range correlation properties are different for protein
coding (exonic) and non coding (intronic, intergenemic) sequences \cite{BUL3}.
One more fundamental ground, there is still continuing debate as to whether
the reported long range correlations really mean a lack of independence at long
distances or simply reflect the patchiness (bias in nucleotide composition) of
DNA sequences. There have been attempts to eliminate local patchiness using
methods such as min-max \cite{PENG1}, detrended fluctuation analysis (DFA) \cite{BUL3,PENG2}
and wavelet analysis \cite{ARN1}. In spite of its success in modeling the long
range correlations observed in DNA sequences, as indicated by the
power law increase
in the variance and the inverse power law spectrum \cite{VOSS,VIE}, the problem of the correct
statistical interpretation of DNA walk is still unresolved and is attracting
the attention of an increasing number of investigators. Since approaches
based on different models predict different correlation structure, there is
no unique measure of the degree of correlation in DNA sequences.
Therefore, it is very important
to investigate the correlations and extract the power law exponent $\alpha$ rather
in a model independent way so that the interpretation of the data including the
theoretical analysis becomes more meaningful.
There is another
confusion related to this study is the absence of a clear definition of the
term "long range". Clearly, what is considered to be long is relative to what
is considered to be short. To over come some of these problems, recently we have
suggested a new method \cite{AKM1} to measure the degree of correlations
using the variance analysis of the density distribution of a single or a group
of nucleotides. We have also suggested a way to find out an approximate length
scale above which all DNA sequences show strong long range correlations irrespective
of their intron contents while below this, the correlation is relatively weak.
Further, the density distribution which is nearly Gaussian at short distances
shows significant deviations  from the Gaussian statistics at large distances.
In this paper, we present the details of
the analyses and also
extract the correlation parameter $\alpha$ for several
intron containing and intronless sequences.

\section {Density distribution and Factorial moments:}
In the present method, we build the frequency spectrum of a
single or a group of nucleotides by dividing the DNA sequence into many
equal intervals of length $l$. For example, to build a purine spectrum,
we compute
\equation n=\sum_{i=l_0}^{l_0+l} u_i \endequation
where $u_i$=1 if the site is occupied by a G or A and $u_i$=0 otherwise.
Ideally, one can divide the entire DNA sequence of length $L$ into $m$
equal intervals of size $l$ $(l=L/m)$. The purine or GA spectrum can be built
by computing $n$ from all the intervals. Alternatively, $n$ can be computed
in any segment between $l_0$ and $l_0+l$ and the spectrum ($n$ distribution
or $P_n$) is built by varying
the starting position $l_0$ from 1, 2, 3 etc upto $L-l$ so as to cover the whole
sequence
\footnote{At short distances, $n$ can be zero
due to the non occurence of a given nucleotide. In such cases, the density
spectrum can be built either including or excluding zero$^{th}$ channel. In this
analysis, we include zero$^{th}$ channel also so that the complementarity is
satisfied which is unlike the case when the zero$^{th}$ channel is excluded.
See appendix B for details}.
We adopt this second procedure for better statistics. Finally, the
standard deviation (SD) of this $P_n$ distribution can be obtained from
$\sigma^2=<n^2-{n_0}^2>$ which in general will depend on the interval or the
window size $l$.

In addition to the standard deviation $\sigma^2$, we also
compute the factorial moments $F_q$'s of $P_n$.
The normalized factorial moments of order q are written as
\equation F_q=\frac{f_q}{f_1^q} \endequation
where
\equation f_q=\sum_{n=q}^{\infty} P_n n(n-1).....(n-q+1)
             =\sum_{n=q} ^{\infty} \frac{n!}{(n-q)!} P_n \endequation
As will be shown later, the factorial moment has the distinct advantage over
the normal moments in identifying the genomic sequence from the random one.
It may be mentioned here that for random  Poisson distribution, the factorial
moments for all q's become unity i.e. for
\equation P_n=\frac{a^n e^{-a}}{n!} \endequation
the above factor for $f_q$ becomes
\equation
f_q=\sum_{n=q}^\infty \frac{n!}{(n-q)!} \frac{a^n e^{-a}}{n!}
   =\sum_{n=q}^\infty \frac{a^n e^{-a}}{(n-q)!}
   =\sum_{m=0}^\infty \frac{a^{m+q} e^{-a}}{m!}
   =a^q\sum_{m=0}^\infty \frac{a^m e^{-a}}{m!}
   =a^q
   \endequation
which gives $F_q$=1.

In this work, we have applied the above factorial moment
analysis (generally used to study the fluctuations during a phase transition
\cite{AKM2}) to study the dynamical fluctuations present in the DNA sequences.

\section {Principle of complimentarity}

\begin{figure}
\centerline{\hbox{
\psfig{figure=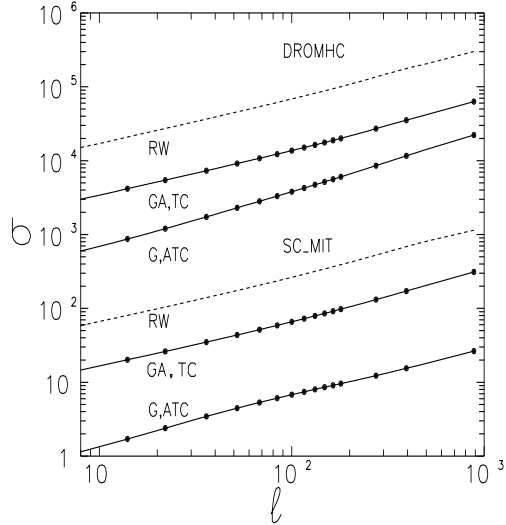,width=3.0in,height=3.2in}}}
\caption{ The variance $\sigma$ versus $l$ for $G$ and $GA$ distributions
(solid curves). Top panel is
for $DROMHC$ (Drosophilia Melanogaster, MHC) while the bottom panel for
  $SC\_MIT$ (yeast mitocondrial DNA). The filled circles are for
  the complimentary $ATC$ and $TC$ distributions.
  The curve $RW$  (dotted curve) corresponds to the
  slope in case of random walk (see text for details). The curves are scaled up appropriately for
  better clarity.}
\label{sd1}
\end{figure}

A general property noticed for all the genomic sequences (of statistically
significant length) with a few exceptions is that the distributions of any
single or group of nucleotides which has a probability of occurrence $p$ has
the same variance $\sigma$ as that of its complimentary group that has the
probability of occurrence $(1-p)$, although both have different distribution
functions. This would imply that even a single nucleotide distribution
say $G$ distribution will have same variance as that of $ATC$ distribution or
a $GA$ distribution will have identical variance as that of $TC$ distribution.
Figure \ref{sd1} shows $\sigma$ versus $l$ plots for $G$ and $GA$ distributions
(solid curves) for two typical sequences of $DROMHC$ (Drosphilia Melanogaster,
MHC, 22663 bps, $20.5 \%$ $G$, $30.3 \%$ $A$, $25.4 \% $ $T$, $23.8 \%$ $C$) and
$SC\_MIT$
(yeast mitochondrial DNA, $9.1 \%$ $G$, $42.2 \%$ $A$, $40.7 \%$ $T$,
$8.0 \%$ $C$).
As can be seen from the figure, the $G$ and $GA$ distributions have same $\sigma$
at all scale as that of $ATC$ and $TC$ distributions (filled circles) although
the distribution functions of the two complimentary groups need not be identical.
The above agreement is exact for most of the DNA sequences
(with a few exceptions) as well as for the
random sequences. For example, the $\sigma$ for
$G$ and $ATC$ distributions of $SC\_MIT$ and $E. Coli:TN10$ ($E. Coli$ with a
$TN10$ mobile transposion (9147 bps) at location 22000 bps) show $2\%$ to $3\%$
deviations at all scale depending on the total
length of the sequences where as for other
DNA as well as random sequences, this
agreement is exact.
(This difference is not visible from figure \ref{sd1}
in case of $SC\_MIT$ as the deviation is insignificant over a large distance).

Within the present formalism, we can also reproduce the result of random walk
$(RW)$ model (See appendix for more detail) by assigning
$u_i=1$ for purine group ($G$ and $A$)
and $u_i=-1$ for pyrimidine group ($T$ and $C$). However, unlike the random
walk model of interpreting $+1$ and $-1$ as the probability of step up and
step down, $P_n$ can be considered as the frequency distribution of $n$
which gives the excess or deficit of purines over pyrimidines. The $\sigma$
versus $l$ as obtained from this assignment has also been shown in
figure \ref{sd1} (see the dotted curves labeled $RW$) for comparison. It is
interesting to note that the $RW$ curves shows a parallel shift with respect
to the $GA$ or $TC$ curves indicating that $GA$ or $TC$ distributions and $RW$
model have similar fluctuations at all scale. This is an interesting
observations, as we can now use $GA$ or $TC$ distributions as alternatives
to the DNA walk representation to study the correlation. The advantage is, since
$n$ represents a sum, unlike the DNA walk model, the entire spectrum lies
to the positive side of the coordinates which is essential to compute various
higher moments like $F_q$ of the distributions.

\begin{figure}
\centerline{\hbox{
\psfig{figure=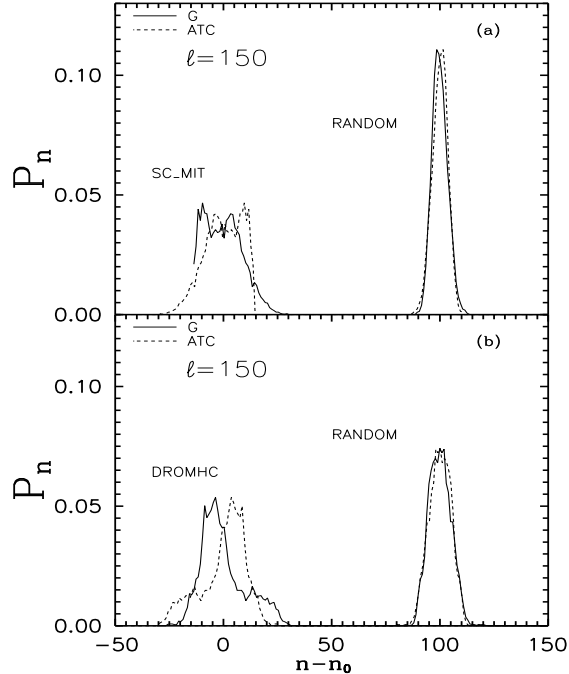,width=4.0in,height=4.2in}}}
\caption{ The complimentary $G$ and $ATC$ density distributions at
a typical distance of $l=150$
for above two sequences. The curves on the right
(shifted by $100$ units) shows the corresponding
distributions in case of a purely random sequence of appropriate $G$, $A$, $T$
and $C$ contents.}
\label{sd2}
\end{figure}

It is also important to note that although the complimentary distributions
have same $\sigma$ at all scale, the distribution functions need not be
exactly identical.
Figure \ref{sd2} shows a typical normalized density distribution functions $P_n$
of two complimentary distributions $G$ and $ATC$
for the above two sequences ($SC\_MIT$ and $DROMHC$)
as a function of $n-n_0$ (where $n_0$ is the average
count ) at a typical length scale of
$l=150$ (figures in left). The figures to the
right shows $P_n$ distributions ($x$-axis is shifted by 100 for clarity)
corresponding to the
two purely random sequences having same length and nucleotide
contents as that of $DROMHC$ and $SC\_MIT$ sequences.
It is interesting to note that although $\sigma$ versus $l$ plots are (nearly)
identical $i. e.$, both distributions have same fluctuations at all scales,
the distribution functions are not identical.
This is an important characteristic of
a DNA sequence which is not found in case of a random one.

\section {Extraction of slope parameter}
\begin{figure}
\centerline{\hbox{
\psfig{figure=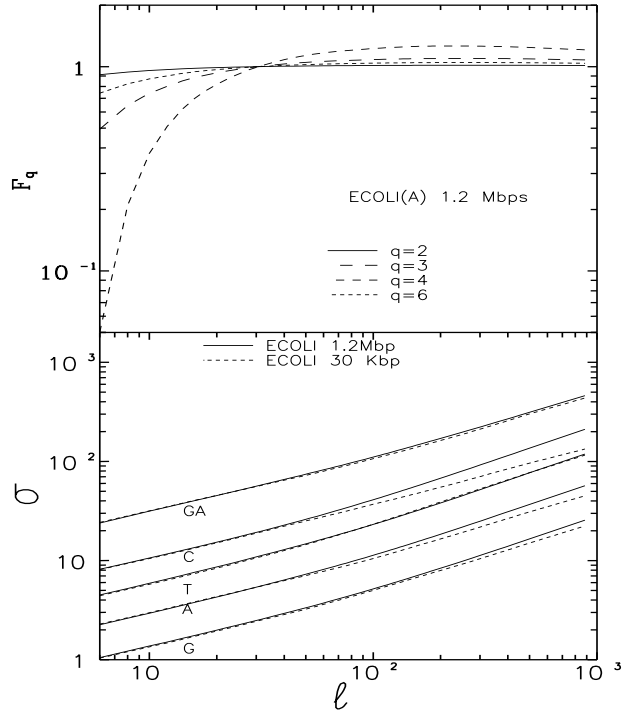,width=4.0in,height=4.2in}}}
\caption{ (a) The factorial moments $F_q$ versus $l$ for a typical $A$ distributions
of $E. Coli$ sequence of length 1.2 Mbps. (b) The corresponding
slope parameter $\sigma$ versus $l$ for $E. Coli$ of length 1.2 Mbps (solid curves)
and of length 30 Kbps (dashed curves). The curves are scaled up appropriately for clarity.}
\label{ec1}
\end{figure}

The long range correlations are generally studied from the relation
$\sigma \sim l^\alpha$ where the parameter $\alpha$ is extracted from the
$\sigma$ versus $l$ plot in the log-log scale. For the case of a completely
random sequence, $\alpha \sim 0.5$. The deviation of $\alpha$ from $0.5$
indicates presence of long range correlations. We have estimated $\sigma$
of $G$, $A$, $T$, $C$ and $GA$ distributions for several DNA sequences and
found that $\sigma$ versus $l$ plot in the log-log scale is not linear over
the entire length \footnote{We consider only the $G$, $A$, $T$ and $C$
distributions to extract the correlation parameters for the individual nucelotides
and $GA$ distributions to simulate the results of random walk model}.
Figure \ref{ec1} shows $\sigma$ versus $l$ plot (bottom panel)
for a typical $E. Coli$ sequence of length $L=1.2$ Mbps (solid curves)
and $L=30$ Kbps (dotted curves) respectively. The top panel shows the factorial
distributions of $q$=2, 3, 4 and 6 for a typical $A$ distributions, although
similar plots can be obtained for other nucleotide distributions as well.
A general feature of the factorial moments of the DNA sequence with a few
exception is that at short distances, $F_q < 1.0$ for all $q's$ and exceeds
unity at some point say at $l_q$. This behavior is not found in case of a purely
random sequence where $F_q$ is always $\le 1.0$. Further, all $q$'s do not
cross unity exactly at the same point, $l_q$ being more for higher $q$ values.
However, this variation is insignificant over a very large scale if we
restrict to some of the lower moments say up to $q=6$.

From these plots and also from the several other studies,
we make following few observations; (i) The $\sigma$ versus $l$ plot is
not linear through out, rather starts bending around some region (say $l_c$,
which could be different for different distributions) indicating a change
of slope from $\alpha_1$ to $\alpha_2$, (ii) For most of the cases, while
$\alpha_1$ shows weak deviation from $0.5$, $\alpha_2$ deviates significantly
from $0.5$ and also depends on the sequence length $L$, (iii) The individual
nucleotide distributions may have stronger correlations than any sum like $GA$
and $TC$ distributions or any other combinations.

\begin{figure}
\centerline{\hbox{
\psfig{figure=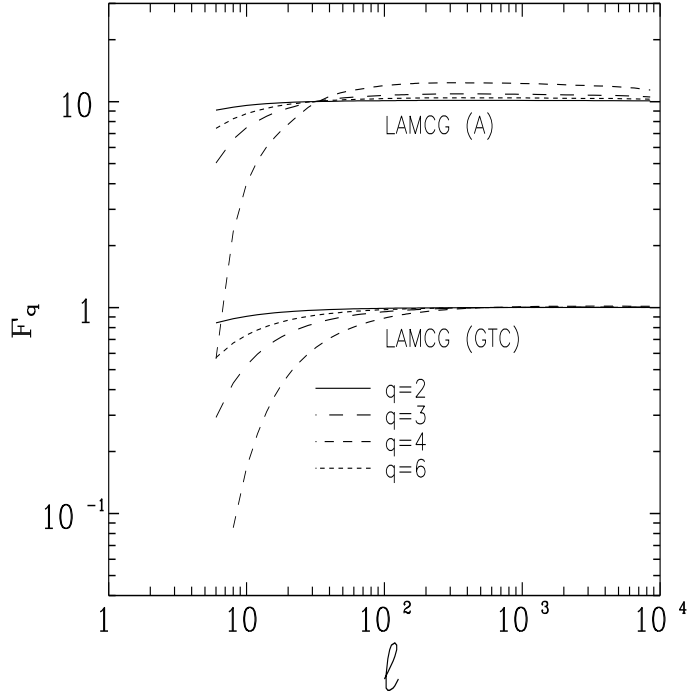,width=4.0in,height=4.2in}}}
\caption{ The factorial moments $F_q$ versus $l$ for $G$ and $ATC$ distributions
of $LAMCG$ sequence}
\label{lam}
\end{figure}

Since $\sigma$ versus $l$ in the log-log plot starts bending around $l_c$,
we can extract the slope by dividing the entire length into two segments;
one for $l<l_c$ and the other one for $l>l_c$. This can be done by examining
each case individually.
However, we have noticed an approximate correlation
between this bending region in $\sigma$ versus $l$ plot
and the cross over
points $l_q$ of the corresponding factorial moments i.e. the slope changes
around the same region where the factorial moments become unity. This
is a pure phenomenological observation which is found for several DNA sequences as listed in tables with a
few exceptions which we will discuss below.
It may be mentioned here that although, the two complimentary distributions
have same fluctuations, both need not have identical factorial moments.
Figure \ref{lam} shows the plots of $F_q$ versus $l$ for $A$ and $GTC$ distribution
for a $LAMCG$ sequence.
Since both are complimentary, they have
identical fluctuations at all scales (hence same bending region), but the
cross over regions in $F_q$ plots are different, being higher for $ATC$ distributions
(due to large average values $n_0$ at all scales). While the $l_q$ value of the
$A$ distribution shows an approximate correlation with the bending region of
$\sigma$ versus $l$ plot where a possible slope change occurs, the $l_q$
values of $GTC$ distribution has no such correlations. This is true for any
complementary distributions of $G$, $A$, $T$ and $C$ except for $GA$ and $TC$ distributions since
both have nearly
same overlapping cross over regions.

\begin{figure}
\centerline{\hbox{
\psfig{figure=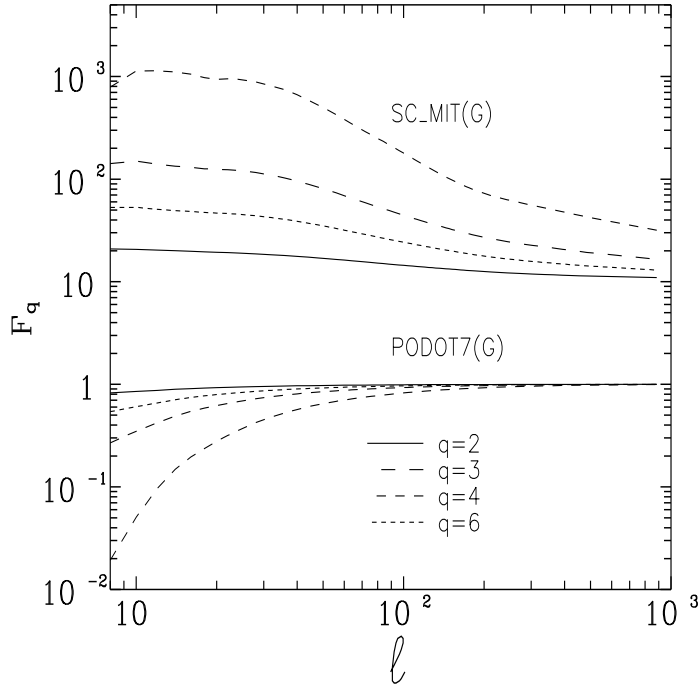,width=4.0in,height=4.2in}}}
\caption{ The factorial moments $F_q$ versus $l$ for $G$ distributions
of $SC\_MIT$ (scaled up) and PODOT7 (T7 bacteriophage) sequences.}
\label{pc1}
\end{figure}

Therefore, only the $l_q$ values of the $G$, $A$, $T$,
$C$ and $GA$ distributions are used as an approximate length scales $(l_c)$.
The entire length of the sequence is divided into
two parts one for $0< l <l_{c1}$ and other for $l_{c2}<l<L_{max}$ where $l_{c1}$
and $l_{c2}$ are the minimum and maximum of all the $l_c$ corresponding to
$G$, $A$, $T$, $C$ and $GA$ distributions. The $L_{max}=L/30$, i.e. we have at
least $30$ independent data sets so that the statistical analysis becomes
meaningful. Therefore, excluding the region $l_{c1}<l<l_{c2}$, we have extracted
$\alpha_1$ and $\alpha_2$ since the linearity in these two segments
are found to be extremely good for most
of the cases.  The results are summarized in three tables which covers
both intronless and
intron containing sequences. The table shows the length of the sequence $L$
used in the analyses, the cross over values $l_q$ ( same as $l_c$),
the slope parameters $\alpha_1$
and $\alpha_2$ and also the corresponding percentage of the nucleotide contents
$P$. A general observation is that the sequence is weakly
correlated at short distance with $\alpha_1$ which is quite close to $0.5$ where as
for $l>l_c$, the correlation is relatively stronger with a larger value
of $\alpha_2$. Now we discuss a few exceptions like in the case of $SC\_MIT$ and
$PODOT7$ ($T7$ bacteriophage, $39936$ bps). Figure \ref{pc1} shows the
factorial moments of a typical $G$ distributions. In both the cases, the factorial
moments do not have any cross over point.
In case of $SC\_MIT$, the factorial moments are much higher than unity
even at small distance and starts decreasing afterwards. The similar behavior
is found for $C$ distribution also. However, the $A$, $T$ and $GA$ distributions
do have $l_c$ points. Therefore, using $l_{c1}$ as $\sim 36$ and $l_{c2} \sim 184$,
we estimated $\alpha_1$ and $\alpha_2$ for $G$, $A$, $T$, $C$ and $GA$ distributions
which are listed in table III. The symbol $'*'$ indicates absence of any critical
value. It is interesting to note that $\alpha_1$ is quite large
and in some cases $\alpha_1 > \alpha_2$.
On the other hand , the factorial moments of the sequence like $PODOT7$ do not
reach unity at any scale. The absence of such type of scale has been indicated by
the symbol $'-'$ in table III. This type of sequences behave like a pure random
one having $\alpha$ values quite close to $0.5$. We have listed a few such sequences
with exceptions in table III.

Further, we would like to mention here that we have noticed that
the factorial moments for many sequences starts decreasing at large distances.
Also for a few cases, the
factorial moments start decreasing even at a very short distances.
Consequently, the slope also changes accordingly. However, we would not
like to assign any reasons due to lack of enough statistics.

The slope with $\alpha=0.5$ corresponds to the case of a normal diffusion
process of a random Brownian trajectory. The basic idea of a Brownian motion is that
of a random walk having a Gaussian distribution probability for the position
of the random walker after a time $t$ with the variance ($\sigma^2$)
proportional to $t$ ($\sigma \sim t^\alpha$ where $\alpha=0.5$).
This corresponds to the case of normal diffusion. However, nature shows
enough examples of anomalous diffusion characterized by a variance
which does not follow a linear growth in time \cite{KLA}.
In such cases either the diffusion is accelerated if $\alpha > 0.5$ or
the growth is
dispersive if $\alpha < 0.5$. As found in the analyses (see tables I and II),
$\alpha_2 > 0.5$ at large distances for most of the sequences irrespective of
their intron contents. However, a few sequences as shown in table III,
not only peculiar, may also have $\alpha$ which decreases at large distances.
In such cases, $\alpha<0.5$ which may indicate the influence of  dispersive
dynamics. This aspect needs further investigations.
Finally, we would like to add here that $\alpha_1$ is close to $0.5$ for
most of the sequences at short distance (see tables I and II). Although, $\alpha=0.5$
would imply about a random behavior, it can not be told conclusively from the
present analyses unless the short distance effects are taken into consideration
\cite{GAL}.

\section{Patchy sequences}

\begin{figure}
\centerline{\hbox{
\psfig{figure=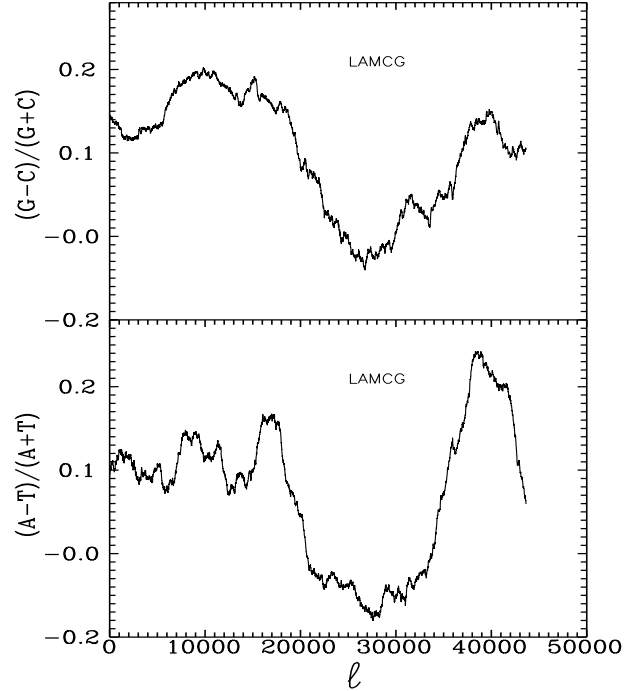,width=4.0in,height=4.2in}}}
\caption{ The $GC$ and $AT$ skews as a function of $l_0$ for $LAMCG$ sequence.}
\label{skew}
\end{figure}

In the following, we investigate whether the mosaic character of DNA
consisting of patches of different composition can account for apparent
long range correlations in DNA sequences\cite{KAR}. The Chargaff's second parity
rule states that in a single strand $G \approx C$ and $T \approx  A$.
However, asymmetries in base composition have been observed in many
sequences. A quantitative estimate of the $GC$ and $AT$ skews  can be
obtained from the relation $(G-C)/(G+C)$ (Excess of $G$ nucleotides over $C$
nucleotides) and $(A-T)/(A+T)$ (Excess of $A$ nucleotides over $T$ nucleotides).
This is, operationally equivalent to estimating $n$ as defined in Eq.(1) except
$n$ now represents the count $(G-C)/(G+C)$
for $GC$ skew and $(A-T)/(A+T)$
for $AT$ skew in a fixed window size of
$(L/20)$. We consider $LAMCG$ as an example and plot $n$ (defined appropriately)
versus $l_0$ where
the starting position of the sliding window $l_0$ varies from $1$, $2$, $3$ etc
upto $L-l$. Figure \ref{skew} shows the plots of $GC$ and $AT$ skews as a function
of the length for a typical $LAMCG$ sequence.
The plots show  a change in the direction of the slope with a change in sign of
the skew. The quantity and quality of the skew can be assessed from the $V$
or from the inverted-$V$ shape of the curves.

From the above plots, we can identify
three well known
compositional domains of $LAMCG$ of size 22000 bps ($GA$ contents 0.54), 17000
bps ($GA$ contents 0.47) and 9000 bps ($GA$ contents 0.54). We also consider
an artificially generated sequence by joining three random
patches of size 22000 bps, 17000 bps and 9000 bps respectively with appropriate
$G$, $A$, $T$ and $C$ contents. We also consider another heterogeneous sequence
generated from $E. Coli$ DNA by
a  mobile insertion of TN10 at location 22000 bps. The corresponding
random patches are of size 22000 bps, 9147 bps and 22000 bps respectively
\footnote{ Please note the distinction between the random sequence
which is generated by joining three random patches of total length $L$
and a pure random one of length $L$. Although, both the sequence has same
percentage of nucleotide contents in the length $L$,
the former is random only patch wise.}

\begin{figure}
\centerline{\hbox{
\psfig{figure=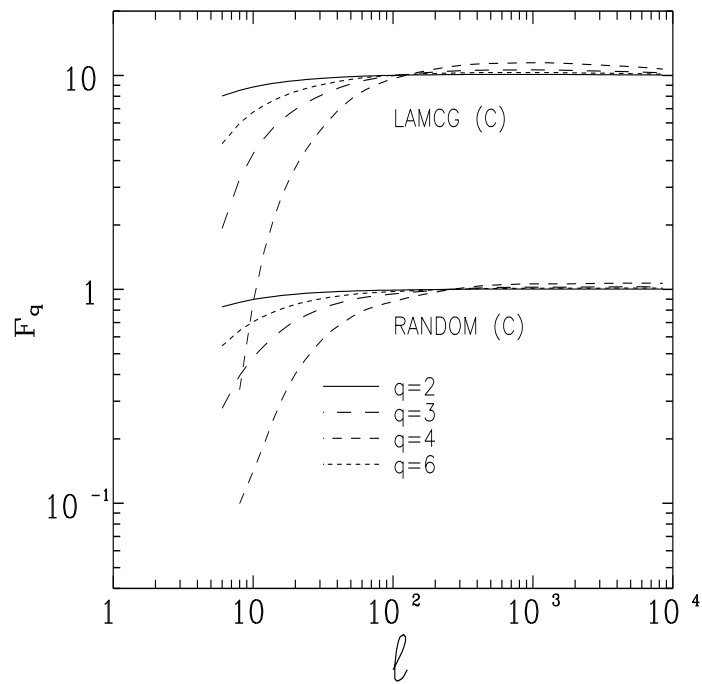,width=4.0in,height=4.2in}}}
\caption{ The $F_q$ versus $l$ of $C$ distribution of
for $LAMCG$ and an artificially
sequence generated by joining three randomly generated patches
of size 22000 bps, 17000 bps and 9000 bps with the same $G$, $A$, $T$ and $C$
contents as that of $LAMCG$.}
\label{lar}
\end{figure}

\begin{figure}
\centerline{\hbox{
\psfig{figure=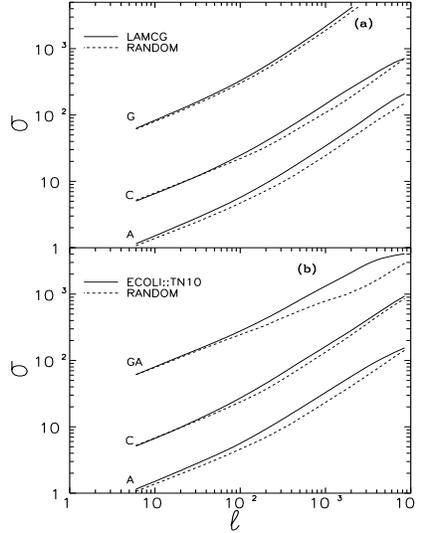,width=3.0in,height=3.2in}}}
\caption{ The variance $\sigma$ versus $l$ for $G$, $A$, $T$, $C$, and $GA$
distributions. (a)  $LAMCG$ and an artificial
sequence generated by joining three randomly generated patches
of size 22000 bps, 17000 bps and 9000 bps with the same $G$, $A$, $T$ and $C$
contents as that of $LAMCG$. (b) for $E. Coli$ with a $TN10$ mobile
transposition (9147 bps) at location 22000 bps. The three random patches
are of size 22000 bps, 9147 bps and 22000 bps with appropriate
$G$, $A$, $T$ and $C$ contents. }
\label{rans1}

\end{figure}

Figure \ref{lar} shows the $F_q$ versus $l$ plot of a typical $C$ distribution
for $LAMCG$ and for an artificially generated sequence (random only patch wise).
Interestingly, the factorial
moments for both the cases behave similarly.
Figure \ref{rans1} shows a similar $\sigma(l)$ versus $l$ plot both for real
and artificially  generated (from random patches) sequences.
Although, in some cases both agree, in general they are not identical at the
individual nucleotide levels particularly at large distances (Note that
the scale is highly compressed). This deviation
would mean that at large distances, the density distribution functions will
have significant discrepancy due to different widths.
So  at a first look from the $\sigma$ versus $l$ plot, we can say that
the actual DNA sequences and the RANDOM patches need not have identical
slopes $\alpha$ (hence the width $\sigma$) at large distances for all
the nucelotides although they agree in some cases.
Even at short distances, although the DNA and the
RANDOM
sequences have nearly identical width $\sigma$, the full shape
of the distributions need
not be identical. To demonstrate this, we invoke the principle of
complimentary which was mentioned before.

\begin{figure}
\centerline{\hbox{
\psfig{figure=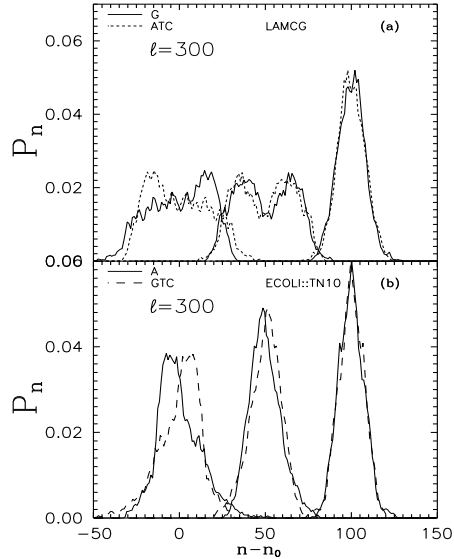,width=3.0in,height=3.2in}}}
\caption{The density distribution $P_n$ versus $n-n_0$ (where
$n_0$ is average density) for a real DNA sequence (left most),
for an artificially generated sequence (middle) and for a completely
random sequence (right most) shown for two complementary
distributions. (a) for $LAMCG$ and (b) for $E. Coli::TN10$.}

\label{fig5}

\end{figure}

Figure \ref{fig5}(a)
shows a $G$ and $ATC$ distribution (left most) for a $LAMCG$ sequence at
$l=300$. Notice that
although $\sigma$ versus $l$ plots are identical, i.e. both distributions have same
fluctuations at all scales, the distribution functions are not same. Such
differences are not found for a real random sequence (right most). The middle
figure corresponds to the case of artificially generated random
sequence. Although, the artificially
generated sequence mimics the real sequence to some extent, it is not fully
capable of reproducing the characteristic of a real sequence. Figure
\ref{fig5}(b)
shows another comparison for a $E. Coli::TN10$ sequence for $A$ and $GTC$
distributions. This discrepancy will be more
prominent at higher $l$ values which the artificially generated sequence can not
reproduce.

\section {Density distributions}

\begin{figure}
\centerline{\hbox{
\psfig{figure=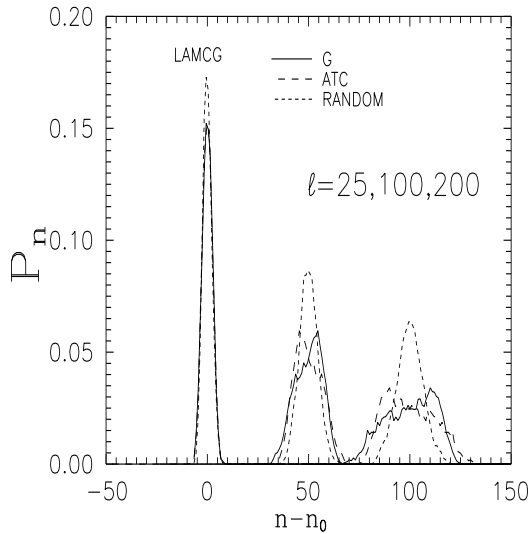,width=3.0in,height=3.2in}}}
\caption{The density distribution $P_n$ versus $n-n_0$ (where
$n_0$ is average density) for $LAMCG$ sequence at $l=25$, $100$
and $200$ respectively. The solid and the dashed curves are for $G$ and
$ATC$ distributions respectively where as the dotted curve is for a
purely random sequence.}
\label{den}
\end{figure}

In \cite{AKM1}, we had demonstrated that the density distribution $P_n$
is Gaussian at short distances and starts deviating from it as the distance
increases. Figure \ref{den} shows another example where $P_n$ has been
plotted for two complimentary distributions at $l=25$, $100$ and $200$ respectively.
The complimentary distributions are nearly identical at short
distance and coincide with the random distributions where as $P_n$ distributions
for $G$, $ATC$ and pure random one are all different at larger distances.

Thus, irrespective of intron contents, most of the sequences follow Gaussian
statistics at short distances. However, at large distances, the statistics
deviates  significantly from the Gaussian nature.

\section {Conclusions}
In conclusion, we have extended our previous work to extract the slope
parameter $\alpha$ for several intron containing and intron less DNA sequences.
The advantage of the present method is that the variance analysis
can be applied to any individual or group of nucleotides. We believe that the
individual nucleotides provide a more fundamental measure of the correlation
than any combination or group (like the DNA walk representation) where the
effects may get reduced or washed out. Another interesting aspect is
the (lower) factorial moments of most of the DNA sequences cross unity in
a very narrow region in $l$ where the $\sigma$ versus $l$ plot in the log-log scale
also shows a bending. Although, a formal justification to this correlation
has not been provided, we have used this scale as an approximate measure
to exclude the bending regions from the slope analyses. Based on this scale,
we divide the DNA sequence into two segments to extract the slope parameters.
It is found that below this scale, the correlation is weak and the DNA
statistics is essentially Gaussian while above this all DNA sequences show
strong long range correlations irrespective of their intron contents with a
significant deviation from the Gaussian behavior. It may be mentioned here
that the controversies that exist in this field of research are primarily
due to different approaches that are adopted in various models. In this context,
our analyses is model independent as it only involves the counting of an individual
or a group of nucleotides in a given length to build the density distribution.
In this work, we do not advocate for any specific model,
although the extracted slope parameters indicate the presence of anomalous
diffusion of both enhanced and dispersive nature. Instead, we
provide an elegant tool to measure
the degree of correlations unambiguously so that the interpretation of
the data including theoretical analyses will become more meaningful. This work will
also provide further impetus to develop models for the understanding of
the DNA dynamics.

\begin{table}
\squeezetable
\caption{Summary of the correlation analysis of intron containing sequences.
$l_c$ is the characteristic length scale.
$\alpha_1$ is the slope parameter for $l<l_{c1}$ and $\alpha_2$ is the slope parameter for
$l_{c2} < l < l_{max}$, where $l_{c1}$ and $l_{c2}$ are the minimum and the
maximum of all the $l_c$, $l_{max}$=L/30 where L is the total length of the
sequence. The acronym in column 1 is the name of the GenBank. Since
the factorial moments for all $q$ do not cross exactly at same point,
we have chosen $l_c$ for which $F_q$ for $q=2,3,4$ and $6$  approaches unity
simultaneously. $P$ denotes percentage of $G$, $A$, $T$ and $C$ in the sequence.
We have also not fine tuned the cross over point $l_c$, it is only approximate.}

\begin{tabular}{|c|c|c| c |c| c| c|c|}

Sequence& L& $l_c$, $\alpha$ &G&A&T&C&GA\\
\hline
Human $\beta$-globin & 73,308  & $l_c$    & 12   & 14   & 14   & 14   & 32\\
(Chromosomal region) &          & $ \alpha_1$   &0.640 &0.644 &0.671 &0.620 &0.652\\
HUMHBB               &          & $\alpha_2$ &0.703 &0.783 &0.812 &0.655 &0.758\\
                     &          &P           &20.2  & 30.1 &30.4 & 19.3 & 50.3\\
\hline
Adenovirus type 2    & 35,937  & $l_c$    & 24   & 12   & 12   & 36   &132\\
(Intron containing)   &         &$\alpha_1$     &0.598 &0.586 &0.567 &0.583 &0.564\\
ADRCG                &         &$\alpha_2$     &0.862 &0.815 &0.816 &0.758 &0.661\\
                     &          &P           &27.3  & 23.2 &21.6 & 27.9 & 50.5\\
\hline
Chicken embryonic MHC& 31,111  &$l_c$     & 24   & 36   &  14   & 28   &48\\
(Gene)               &      &$\alpha_1$        &0.644 &0.578 &0.658  &0.581 &0.623\\
CHKMYHE              &      &$\alpha_2$        &0.775 &0.698 &0.800   &0.715 &0.762\\
                     &          &P           &22.2  & 31.3 &26.7 & 19.8 & 53.5\\
\hline
Human $\beta$-cardiac MHC& 28,438  &$l_c$     & 16   & 16   &  10   & 18   &20\\
(Gene)               &      &$\alpha_1$        &0.638 &0.579 &0.627  &0.620 &0.664\\
HUMBMYH7              &      &$\alpha_2$        &0.681 &0.663 &0.700   &0.673 &0.688\\
                     &          &P           &25.9  & 23.6 &23.0 & 27.5 & 49.5\\
\hline
Drosophila melanogaster MHC& 22,663  &$l_c$     & 20   & 20   &  14   & 36   &156\\
(Gene)               &      &$\alpha_1$        &0.648 &0.594 &0.644  &0.562 &0.569\\
DROMHC                     &      &$\alpha_2$        &0.820  &0.652 &0.798  &0.707 &0.719\\
                     &          &P           &20.5  & 30.3 &25.4 & 23.8 & 50.8\\
\hline
Chicken c-myb oncogene    & 8200  &$l_c$& 14  & 10   &  10   & 12   &48\\
(Gene)              &    &$\alpha_1$      &0.663 &0.661 &0.688  &0.670 &0.645\\
CHKMYB15            &    &$\alpha_2$      &0.749 &0.873 &0.752  &0.852 &0.550\\
                     &          &P           &28.4  & 21.9 &23.5 & 22.2 & 50.3\\

\end{tabular}
\end{table}

\begin{table}

\caption{Same as table I, but for intron less sequences.
For $E. Coli$,
$l_{max}$ is chosen as 120,0000 bps. The data is taken from the site
{\bf http://www.ncbi.nlm.nih.gov}.}

\begin{tabular}{|c|c|c| c |c| c| c|c|}
Sequence& L& $l_c$, $\alpha$ &G&A&T&C&GA\\
\hline
$E. Coli K12$           & 1200000&$l_c$        & 100  & 32   &  32   & 92   &684\\
                       &    &$\alpha_1$        &0.535 &0.542 &0.549  &0.532 &0.529\\
                       &    &$\alpha_2$        &0.665 &0.639 &0.664  &0.674 &0.614\\
                       &    &$\alpha_2$        &0.654 &0.654 &0.655  &0.715   &0.563\\
                 &    &P                 &27.2  &23.6  &24.2   &25.0  & 50.8\\
\hline
H. Influenzae                    & 240000&$l_c$        &  52  & 48   &  56   & 52   &214\\
                       &    &$\alpha_1$        &0.542 &0.552 &0.543  &0.547 &0.543\\
                       &    &$\alpha_2$        &0.720 &0.712 &0.635  &0.770 &0.709\\
                 &    &P                 &17.9  &31.6  &30.7   &19.8  & 49.5\\
\hline
Bacillus subtilis                  & 3840x60&$l_c$        &  80  & 40   &  22   & 132   &274\\
                       &    &$\alpha_1$        &0.538 &0.545 &0.550  &0.508 &0.536\\
                       &    &$\alpha_2$        &0.815 &0.770 &0.816  &0.779 &0.766\\
                 &    &P                 &24.5  &29.5  &26.5   &19.5  & 54.0\\
\hline
Mycobacterium                 & 9665x60&$l_c$        &  20  & 64   &  44   & 24   &136\\
tuberculosis                       &    &$\alpha_1$        &0.549 &0.535 &0.548  &0.540 &0.542\\
                 &    &$\alpha_2$        &0.827 &0.681 &0.826  &0.765 &0.791\\
                 &    &P                 &15.92  &34.57  &33.73   &15.78  & 50.49\\
\hline
Cyano bacterium                   & 4166x60&$l_c$     &  32  & 40   &  28   & 24   &304\\
                       &    &$\alpha_1$        &0.545 &0.532 &0.542  &0.541 &0.535\\
                       &    &$\alpha_2$        &0.730 &0.678 &0.763  &0.733 &0.587\\
                 &    &P                 &24.1  &26.0  &26.0   &23.9  & 50.1\\
\hline
Schizosaccharomyces    & 19431      &$l_c$     & 32   & 60   & 80    &304  &160\\
Mitochondiron          &    &$\alpha_1$        &0.547 &0.561 &0.568  &0.504  &0.543\\
NC-001326              &    &$\alpha_2$        &0.698 &0.690 &0.774  &0.465  &0.773\\
                       &    & P                &15.8  &33.8  &36.1   &14.3   &49.6 \\
\hline
Human Cytomegalovirus  & 229354 &$l_c$     & 36   & 10   & 10    & 32   &148\\
Strain AD169                &    &$\alpha_1$        &0.582 &0.588 &0.596  &0.581 &0.575\\
HEHCMVCG                    &    &$\alpha_2$        &0.806 &0.799 &0.800   &0.800  &0.682\\
\hline
dmal                   &889x60&$l_c$      & 20   & 12   & 12    & 22   &68\\
                       &    &$\alpha_1$        &0.575 &0.628 &0.599  &0.559 &0.60\\
                       &    &$\alpha_2$        &0.730 &0.782 &0.602  &0.720 &0.596\\
\hline
Chicken nonmuscle MHC       &7003  &$l_c$     & 96  & 72   & 12   & 28   &  64\\
(cDNA)                 &    &$\alpha_1$        &0.573 &0.538 &0.569  &0.554    &0.627\\
CHKMYHN                &    &$\alpha_2$        &0.722 &0.833  &0.841  &0.601   &0.842\\
                 &    &P                 &27.0  &31.2  &20.6   &21.2  & 58.2\\
\hline
Bacteriophage $\lambda$& 48,502&$l_c$     & 56   & 36   &  18   &124   &168\\
(Intronless virus)     &    &$\alpha_1$        &0.563 &0.541 &0.598  &0.513 &0.550\\
LAMCG                  &    &$\alpha_2$        &0.935 &0.819 &0.911  &0.810 &0.866\\
                 &    &P                 &26.4  &25.4  &24.7   &23.5  & 51.8\\
\hline
Human dystrophin       & 13,957&$l_c$     & 136  & 56   &  14   & 22   &128\\
(cDNA)                 &    &$\alpha_1$        &0.530 &0.552 &0.569  &0.552 &0.544\\
HUMDYS:M18533          &    &$\alpha_2$        &0.738 &0.634 &0.777  &0.720 &0.725\\
                       &    &     P            &22.4  &33.0  & 24.7  &19.9  &55.4\\

\end{tabular}
\end{table}

\begin{table}

\caption{Same as table II.
The symbol $*$ indicates that the factorial moments are larger
than unity even at very short distance where as $-$ indicates that the factorial
moments do not reach unity.}

\begin{tabular}{|c|c|c| c |c| c| c|c|}
Sequence& L& $l_c$, $\alpha$ &G&A&T&C&GA\\
\hline
SC-MIT                 & 85779 &$l_c$     & *  & 36   & 36    & *  &184\\
Nc-001224              &    &$\alpha_1$        &0.732 &0.697 &0.680  &0.720  &0.578\\
                       &    &$\alpha_2$        &0.698 &0.540 &0.747  &0.508  &0.730\\
                       &    & P                &9.1   &42.2  &40.7   &8.0    &51.3 \\
\hline
Pichia canadensis      & 27694      &$l_c$     & *    & 36   & 64    &*  &96\\
Mitochondiron          &    &$\alpha_1$        &0.654 &0.688 &0.624  &0.615  &0.620\\
NC-001762              &    &$\alpha_2$        &0.662 &0.755 &0.784  &0.660  &0.801\\
                       &    & P                &10.2  &41.6  &40.2   &8.0    &51.84 \\
\hline
Ti(Plasmid)            &24595  &$l_c$     & 76  & 24   & 32    & 40   & -\\
                       &    &$\alpha_1$        &0.543 &0.564 &0.552  &0.586    &0.508\\
                       &    &$\alpha_2$        &0.706 &.700  &0.676  &0.728   &0.433\\
                 &    &P                 &23.5  &26.6  &27.5   &22.4  & 50.1\\
\hline
BacteriophageT7                 &39937  &$l_c$     & -  & 116   & 884    & 1284   &-\\
NC-001604               &    &$\alpha_1<116$        &0.526 &0.571 &0.529  &0.530    &0.530\\
                       &  &$116<\alpha_2<1330$        &0.560 &0.587  &0.590  &0.566   &0.551\\
                 &    &P                 &25.8  &27.2  &24.4   &22.6  & 53.0\\
\hline
Tyorg                  & 196x60 &$l_c$     &  - & 96 &- &36 & 96\\
                       &        &$\alpha_1$&0.491 & 0.560 & 0.515 & 0.620 & 0.587\\
                       &        &$\alpha_2$&0.370 & 0.715 & 0.514 & 0.799 & 0.704\\
                 &    &P                 &16.0  &35.9  &26.7  &21.4  & 51.9\\
\end{tabular}
\end{table}

\appendix
\renewcommand{\thefigure}{A\arabic{figure}}
\section *{Random walk model}

The method of DNA walks, first suggested by Peng et al \cite{PENG1} is based
on the rule that the walker either moves up $(u_i=1)$ or down $u_i=-1)$ for each
step $i$ of the walk. This is the case of a correlated random walk and differs
from an uncorrelated walk where the direction of each step is independent of the
previous steps. Further they assign $u_i=1$ if a pyrimidine occurs at the site
$i$ whereas $u_i=-1$ if the site contains a purine.
The net displacement $(y)$ of the walker after $l$ steps is defined
as
\equation
y(l)=\sum_{i=1}^l u(i)
\endequation
The standard deviation of the above quantity can be estimated from
\equation
\sigma^2(l,L)=\frac{1}{L-l} \sum_{l_0=1}^{L-l} (\Delta y(l_0,l)-{\bar {\Delta(l)}})^2
\endequation
where $L$ is the number of nucleotides in the entire sequence and
\equation
{\bar {\Delta y(l)}}=\frac{1}{L-l} \sum_{l_0=1}^{L-l} \Delta y(l_0,l)
\endequation
where $\Delta y(l_0,l)=y(l_0+l)-y(l_0)$.
It was found \cite{PENG1} that the fluctuations can be approximated by
\equation
\sigma(l,L) \sim l^\alpha
\endequation
where $\alpha$ is the correlation exponents. For $\alpha$ close to $0.5$, there
is no correlation or only short range correlation in the sequence. If $\alpha$
is significantly different from $0.5$, it indicates long range correlations.

\appendix
\setcounter {figure}{0}
\renewcommand{\thefigure}{B\arabic{figure}}
\section *{B}

\begin{figure}
\centerline{\hbox{
\psfig{figure=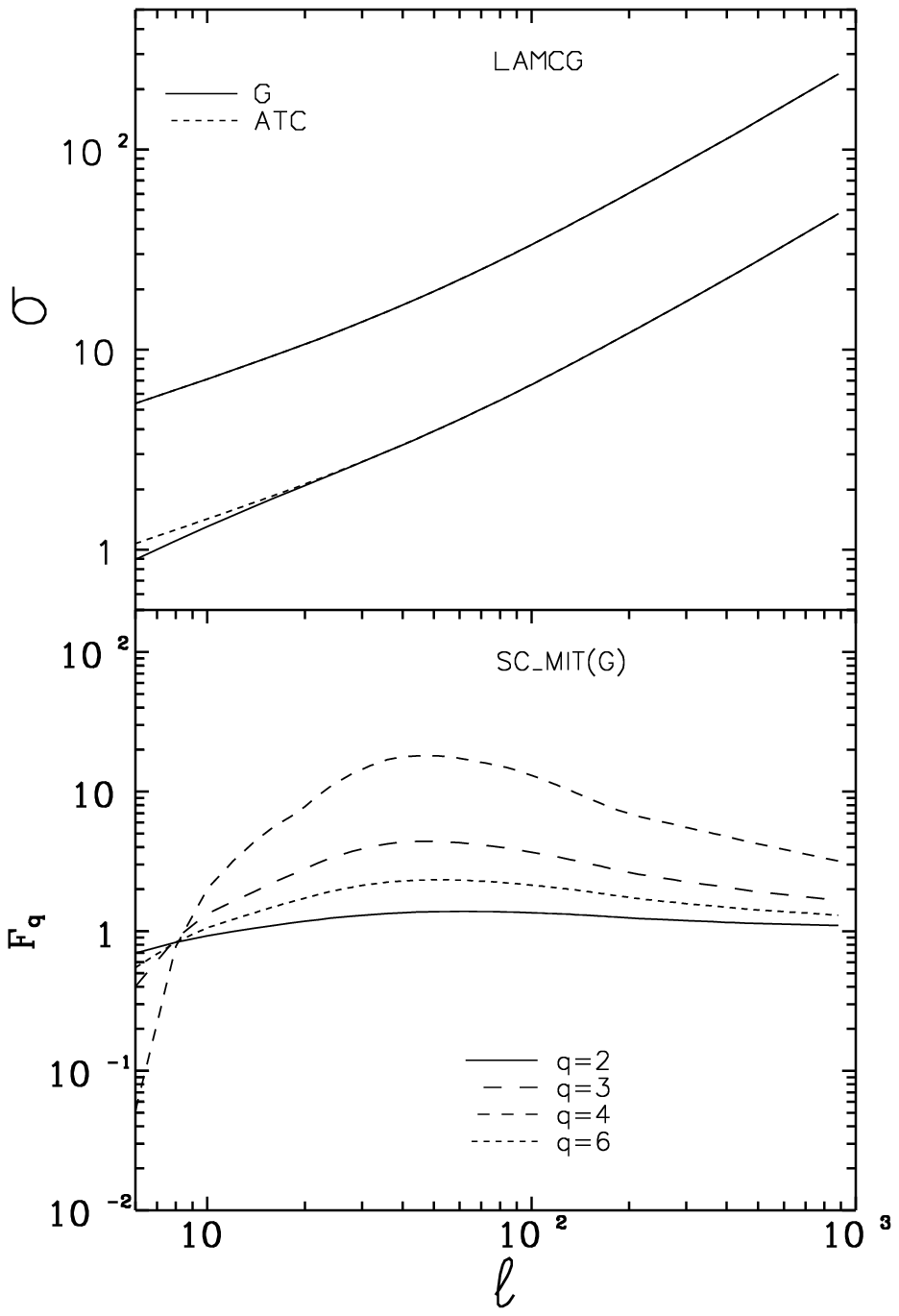,width=3.0in,height=3.2in}}}
\caption{ (a) The variance $\sigma$ versus $l$ for $G$ (solid curves)
and $ATC$ distributions (dotted curves) for $LAMCG$ sequence.
Top panel is
for distribution for which the complimentarity is preserved
while complimentarity is not satisfied in the case of bottom panel particularly
at small distances. (b) $F_q$ versus $l$ plot for $G$ distribution of $SC\_MIT$
for the case when complimentarity is not preserved.
The curves are scaled up appropriately for
 better clarity.}
\label{ap1}
\end{figure}

In the previos analyses, we account for the non-occurence of a particular
nucleotide. This is operationally equivalent to building the density spectrum
$P_n$ including $n=0$. If the nucleotide compositional asymmetry is  quite large like
$SC\_MIT$, the occurence $n$ can be zero for some nucleotides particularly at
short distances. Therefore, we can build $P_n$ distribution either including
or excluding zero$^{th}$ channel. The figure \ref{ap1}(a) shows the comparison
of $\sigma$ versus $l$ plot for two complimentary distributions corresponding
to a $LAMCG$ sequence both with (top panel where $G$ and $ATC$ distributions have
identical slopes at all scales) and without (bottom panel)
inclusion of $n=0$ channel in the $P_n$ spectra. Interestingly, absence of
$n=0$ channel does not satisfy the complimentarity relation particularly at
short distances. However, the difference does not exist at larger distances
where always $n>1$. Figure \ref{ap1}(b) shows another example of $F_q$ versus
$l$ plot for a typical $SC\_MIT$ sequence. The spectrum with exclusion of
$n=0$ channel behaves differently when zero$^{th}$ channel is included (compare
it with figure \ref{pc1} where $F_q$ versus $l$ has no cross over).

Since the spectrum behaves differently when zero$^{th}$ channel is not included,
we have analysed the spectrum of three typical sequences listed in the table below.
Notice now that while $\alpha_2$ values are essentially same as before, the
$\alpha_1$ values are quite different. In fact, we have noticed a general
trend where $\alpha_1$ is higer than the previous values although the corresponding
density distributions do not deviate significantly from the Gaussian behavior
at short distances. However, in the previous analysis, we alwyas include the
zero$^{th}$ channel so that the complimentarity properties is satisfied at all
scales. Moreover, we also found a correlation between $\alpha$ and Gaussian
statistics, namely the deviation of $\alpha$ from $0.5$ also shows a
corresponding deviation of $P_n$ distribution from Gaussian behavior.
For example, in case of $SC\_MIT$, the $\alpha$ is quite large at a short
distance. Accordingly, the $P_n$ distribution also shows strong deviation from
the Gaussian statistics. However, this is not  necessarilly true when
complimentarity is not preserved while building the spectrum.
At short distances, the deviation of $\alpha$ from $0.5$
does not always mean a strong deviation from the Gaussian statistics.

\begin{table}

\caption{The slope parameters for three typical sequences where the complimenraity
is not preserved.}

\begin{tabular}{|c|c|c| c |c| c| c|c|}
Sequence& L& $l_c$, $\alpha$ &G&A&T&C&GA\\
\hline
Bacteriophage $\lambda$& 48,502&$l_c$     & 56   & 36   &  18   &124   &168\\
(Intronless virus)     &    &$\alpha_1$        &0.720 &0.670 &0.740  &0.680 &0.580\\
LAMCG                  &    &$\alpha_2$        &0.935 &0.819 &0.910  &0.800 &0.860\\
                 &    &P                 &26.4  &25.4  &24.7   &23.5  & 51.8\\
\hline
SC-MIT                 & 85779 &$l_c$     & 14  & 36   & 40    & 12  &184\\
Nc-001224              &    &$\alpha_1$        &0.703 &0.760 &0.750  &0.700  &0.630\\
                       &    &$\alpha_2$        &0.694 &0.540 &0.750  &0.510  &0.730\\
                       &    & P                &9.1   &42.2  &40.7   &8.0    &51.3 \\
\hline
BacteriophageT7                 &39937  &$l_c$     & -  & 116   & 884    & 1284   &-\\
NC-001604               &    &$\alpha_1<116$        &0.560 &0.610 &0.570  &0.570    &0.530\\
                       &  &$116<\alpha_2<1330$        &0.560 &0.587  &0.590  &0.566   &0.551\\
                 &    &P                 &25.8  &27.2  &24.4   &22.6  & 53.0\\
\end{tabular}
\end{table}

\end{document}